\documentclass{sig-alternate}

\usepackage{graphicx}
\usepackage{comment}
\usepackage{epstopdf}
\usepackage{subcaption}
\usepackage{color}
\usepackage[ampersand]{easylist}
\usepackage{pgf,tikz}
\usepackage{paralist}
\usetikzlibrary{arrows, positioning}
\usepgflibrary[shapes.misc]
\usepackage{hyperref}
\definecolor{subtler}{rgb}{1,0,0.1}    

\input{notation.tex}

\begin{document}
\title{An Inverse Problem Approach for Content Popularity Estimation}

\numberofauthors{2} 
\author{
	\alignauthor
	Felipe Olmos\\
	\affaddr{Orange Labs and CMAP \'Ecole Polytechnique}\\
	\email{luisfelipe.olmosmarchant@orange.com}
	\alignauthor
	Bruno Kauffmann\\
	\affaddr{Orange Labs}\\
	\email{bruno.kauffmann@orange.com}
}
\maketitle
\begin{abstract}
The Internet increasingly focuses on content, as exemplified by the now popular
Information Centric Networking paradigm. This means, in particular, that
estimating content popularities becomes essential to  manage and distribute
content pieces efficiently. In this paper, we show how to properly estimate
content popularities from a traffic trace.

Specifically, we consider the problem of the popularity inference in order to
tune content-level performance models, e.g. caching models.  In this context,
special care must be brought on the fact that  an observer measures only the
flow of requests, which differs from the model parameters, though both
quantities are related by the model assumptions. Current studies, however,
ignore this difference and use the observed data as model parameters.  In this
paper, we highlight the inverse problem that consists in determining parameters
so that the flow of requests is properly predicted by the model.  We then show
how such an inverse problem can be solved using Maximum Likelihood Estimation.
Based on two large traces from the Orange network and two synthetic datasets,
we eventually quantify the importance of this inversion step for the
performance evaluation accuracy.
\end{abstract}

\category{C.2}{Computer-Communication Networks}{General}

\keywords{Popularity Distribution, Mixture Model, Maximum
Likelihood Estimation, Performance models, Caching} 

\section{Introduction}
``Content is king'', says nowadays a popular Internet meme. This advent of
ubiquitous content is reflected on the Internet, both by the importance of
Content Distribution Networks (CDNs) and transparent caching for coping with an
ever-increasing traffic demand, and by the emergence of the Information Centric
Networking (ICN) paradigm. Understanding content and, in particular, its
popularity is now essential to improve the Internet and its applications.
Content-level performance models are therefore a key tool in the analysis,
design and dimensioning of networks.

Sparse models are particularly useful, since they capture the salient features
of the system while remaining simple enough for analysis, depending only on a
few parameters. These parameters have a large impact on the model output; yet
they often cannot be observed directly in measurements. Carrying a sensible
analysis using the chosen model therefore requires solving the \emph{inverse
	problem} to find the best model parameters of the system from the
measurements.

Due to the rise of content, the number of available documents and their
popularity distribution are now key parameters for traffic models. They have
attracted significant attention from the community in the context of user
generated content  \cite{cha2007tube, gill2007youtube}, HTTP traffic
\cite{gong2001tails,imbrenda2014icn}, and peer-to-peer networks
\cite{clauset2009power, roberts2013exploring}. However, the measurement methods
used in these works are not suited for parameterizing a performance model. In
fact, they fail to take into account that the request count for a given
document in a given observation period, within the framework of a stochastic
model, is not a fixed value, but a random variable. In particular, they ignore
the fact that, in traffic traces, objects with no request are not observed,
being thus a \emph{zero-censored} sample.

Our main objective in this paper is to provide a sound methodology for
popularity estimation, with the aim of correctly fitting performance models.
This requires to take into account the stochastic relation between the model
parameters and the request counts that are observed in a given dataset.  To
this aim, we follow \cite{clauset2009power} in constructing Maximum
Likelihood estimates. We illustrate the aforementioned issues and
methodologies in the case of Poisson based traffic models in the context of
caching performance. Nonetheless, the essential paradigm that we propose is
applicable to other traffic models and contexts. Note that the choice of
relevant performance models is outside the scope of this paper.

The rest of this paper is organized as follows. We first review the literature
in Section~\ref{sec:related}, and describe in Section~\ref{sec:data} the
datasets we use.  We then explicitly identify and formulate in
Section~\ref{sec:background} the inverse problem that consists in correctly
calibrating performance models from trace measurements. To our knowledge, such
a formulation has not been provided in previous studies. In
Section~\ref{sec:est}, we propose a ML estimation method for this inverse
problem. Section~\ref{sec:apps} provides a numerical evaluation of our
approach. Our results and possible extensions are discussed in
Section~\ref{sec:disc}.

\section{Related Work}
\label{sec:related}
The related work we here review falls into two broad categories: content
popularity estimation from measurements and statistical methods.

Due to the fact that popularity distributions usually exhibit a power law
behavior, a common method to estimate them is to fit its rank-frequency
distribution in double logarithmic scale. This approach has been recently
criticized by Clauset et al.~\cite{clauset2009power}. The main issue is that
the rank-frequency plot is not a reliable statistic since, for example, it can
exhibit power-law behavior even if the ground-truth does not.

Despite these problems, the use of the latter method is still pervasive in
performance evaluation~\cite{fricker2012impact, guillemin2013experimental} and
traffic characterization studies~\cite{guo2008stretched, imbrenda2014icn,
	carlinet2012four}. Authors try to  improve  these methods by means of various
adjustments. In~\cite{imbrenda2014icn}, for example, authors separate in three
parts the rank-frequency plot adjusting different curves in each piece and
in~\cite{guo2008stretched}, authors adjust ``stretched exponential'' curves
instead of power-laws.

The latter adjustments indeed solve some of the fitting issues. In previous
studies~\cite{guillemin2013experimental, olmos2014catalog}, we  have noted
another issue in the context of performance models, which arises from the fact
that it is permitted to objects to have zero request. In consequence, from the
point of view of the network operator, objects with no request are not observed
in traces. In the statistical jargon, this is called \emph{zero-censored} and
not taking this fact into account leads one to underestimate the catalog size,
which has an impact on the conclusions drawn from the fitted model (see
Section~\ref{sec:apps}).

In the present work, we address the previous issues by using Maximum Likelihood
(ML) estimates. This method allows us to seamlessly handle the zero-censored
case and it is proposed by Clauset et al.~\cite{clauset2009power} as a robust
method to fit heavy tailed data, which is a common property in popularity
distributions. Maximum likelihood methods have already been in use for flow
size estimation~\cite{loiseau2009maximum} and call center
modeling~\cite{oreshkin2014rate}. The latter work uses an approach similar to
ours, but it is limited to a specific parametric model for non-censored data.
More importantly, our work highlights the fact that the assumptions of the
performance model must be taken into account for a proper popularity
estimation.

The statistical basis of our methods is the estimation of mixed discrete
distributions, a subject that has been extensively studied  in the literature.
The non-parametric case has been addressed from two points of view: the first
one searches the mixing density in the space generated by Laguerre
polynomials with an exponential cut-off; the estimator is then obtained by a
projection on the latter space~\cite{roueff2005mixture, comte2015laguerre}. It,
however, converges slowly with the sample size unless the density belongs to
the aforementioned space.  We therefore base our methodology on the second
point of view, which assumes the mixing distribution to be a sum of Dirac
masses. The estimation methods are then similar to an Expectation-Maximization
scheme (EM)~\cite{lindsay1995mixture}. As regards the parametric case, EM
schemes for finding the parameters of the mixing distribution are provided for
many families in~\cite{karlis2001general}. In both parametric and
non-parametric cases, the estimation algorithms do not handle the case of
censored data, and thus we simply use an all-purpose nonlinear optimization
solver to obtain our results.

\section{Datasets}
\label{sec:data}
We base our analysis on two real-traffic datasets, called \data{yt} and
\data{vod} respectively. Dataset \data{yt} comes from the YouTube traffic
delivered for three months in 2013 by the Orange Network in Tunisia, while
\data{vod} comes from the Video-on-Demand Orange service in France for 3.5
years. The traffic consists in $46\,000\,000$ (resp.  $\!\!3\,400\,000$)
requests to $6\,300\,000$ (resp. $\!\!120\,000$) videos in the \data{yt} (resp.
\data{vod}) set. More details on the collection and processing of these two
datasets can be found in~\cite{olmos2014catalog}.

We also use two synthetic datasets, called \data{prt} and \data{delta}.  This
allows us to highlight in a more clear way some of our findings and, more
importantly, to validate the results with controlled experiments when the
ground-truth is not available.  The set \data{prt} (resp.
\data{delta}) is generated by first drawing $10\,000\,000$ (resp. $100\,000$)
random samples with distribution $\ParetoD{1.6}{0.1}$ (resp. Dirac delta at
$4.0$) representing the popularity (see section \ref{sec:est_mixt} for a model
description).  The number of requests for each document is then drawn according
to the Poisson distribution with mean equal to the document popularity. After
discarding the documents with zero request, this results into $2\,600\,000$
(resp. $400\,000$) requests to $1\,900\,000$ (resp. $98\,000$) documents.

\section{Problem Definition}
\label{sec:background}
In the following, we are given a stochastic object-level model predicting some
performance indicator. The predicted performance explicitly depends on a few
parameters which characterize each object (e.g., document popularities,
lifespans, sizes). It also strongly depends, however, on some implicit
assumptions about the traffic or request process.

An example of such a situation is the evaluation of the hit ratio of a Least
Recently Used (LRU) Cache, which is typically performed using  the Independent
Reference Model (IRM). In this context, users request documents among a catalog
of $\CatSize$ documents. These requests are intercepted by a cache server,
which can store and serve only an evolving subset of the catalog.  The IRM
assumes that the sequence of requests for document $1\leq \iDoc \leq \CatSize$
is a Poisson process with intensity $\Pop_\iDoc$, where $\Pop_\iDoc$ is
proportional to the popularity of document $\iDoc$; all such processes are
mutually independent and their superposition build up the total request
process. In this model, the number $\NReqs_\iDoc$ of requests for document
$\iDoc$ in a time window $\WinSize$ is an independent  Poisson random variable
$\Poiss{\Pop_\iDoc\WinSize}$ of mean $\Pop_\iDoc\WinSize$. Up to a time
normalization, we assume in the following that $\WinSize = 1$.

\begin{figure}
\begin{tikzpicture}
[int/.style={draw,rounded rectangle,minimum width=0.45\columnwidth},
  ed/.style={<-, shorten >=2pt, shorten <=2pt, semithick}]
\node[int] (Params) {Model parameters};
\node[int,below=.4cm of Params] (Trace) {Request Flow}
edge[ed] (Params);
\node[int,below=.4cm of Trace] (Perf) {Performance Indicator}
edge[ed] (Trace);

\node[right=.05\columnwidth of Params] (dummy1) { };
\node[below=of dummy1] (dummy2) { };
\node[below=.7cm of dummy2] (dummy3) { };

\node[int, right=.05\columnwidth of dummy1] (Params2) {$\CatSize,\FSeq{\Pop}{\CatSize}$};
\node[int,below=.3cm of Params2] (Trace2) {$\widetilde{\CatSize}, \FSeq{\NReqsR}{\widetilde{\CatSize}}$}
edge[ed] (Params2);
\node[int,below=.3cm of Trace2] (Perf2) {$C \mapsto \HitRatio(C)$}
edge[ed] (Trace2);

\draw[thick] (dummy1)++(0,0.2) -- (dummy3) ;
\end{tikzpicture}
\vspace*{-.5cm}
\caption{Schematic view of a performance model (left); example
  in the IRM case (right)}
\label{fig:eval_process}
\end{figure}
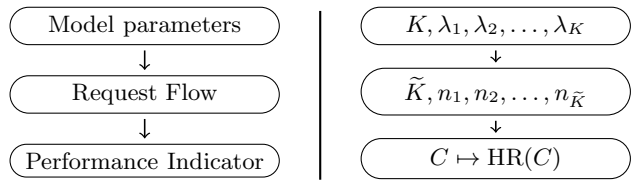

Figure \ref{fig:eval_process} illustrates those two stages, both for an
arbitrary performance model and the IRM case. The first stage consists in
mapping the model parameters to a request flow (or a request flow
distribution). The second step of the model computes the performance indicator,
based on this request flow. In order to keep this paper concise, we now limit
ourselves to the IRM model (see Section ~\ref{sec:extension} for extensions).

Assume now that an observer has access to a sample of the actual request flow,
e.g., a trace dataset or server logs. In the case of IRM, a sufficient
statistics of the request process is the request counts
$\FSeq{\NReqsR}{\ObsDoc}$ for all observed document, where $\ObsDoc$ is the
number of observed documents in the sample.  Following the point of view of an
Internet Service Provider (ISP), we here assume  that objects with zero request
\emph{are not observable} in the sample.  Our main objective  is to solve the
following inverse problem: \emph{estimate the popularity distribution such that
the request flow predicted by the model using these parameters represents the
data at best}.

A simple way to do this, henceforth called the \emph{naive method}, consists in
estimating the popularity of a document by its request count and the catalog
size by the number of observed objects, that is:
$\Est{\CatSize}^{\nv}   = \ObsDoc $ and
$\Est{\Pop}_\iDoc^{\nv} = \NReqsR_\iDoc$,
for $ 1 \leq \iDoc \leq \Est{\CatSize}^{\nv}$.

Two problems can be identified at this stage. First, since the trace is
zero-censored, with high probability the observed number of documents $\ObsDoc$
is strictly smaller than the catalog size $\CatSize$. Second,  each document
popularity $\Pop_\iDoc$ is estimated by a single sample $\NReqsR_\iDoc$ of the
random count $\NReqs_\iDoc$.  This last limitation is well illustrated in the
case of the \data{delta} dataset. By definition, the ground-truth (real)
popularities are $\Pop_\iDoc = 4$. In the dataset, however, the counts of
document requests are Poisson random variables of mean 4, hence
$\Est{\Pop}_\iDoc^{\nv} = \Poiss{4}$ and the naive estimation ``dilutes'' the
mass of popularities over the set of positive integers.
\begin{figure}[b!ht]
  \begin{subfigure}{0.3\textwidth}
    \label{fig:hr_pareto_vs_naive}
    \centering
    \input{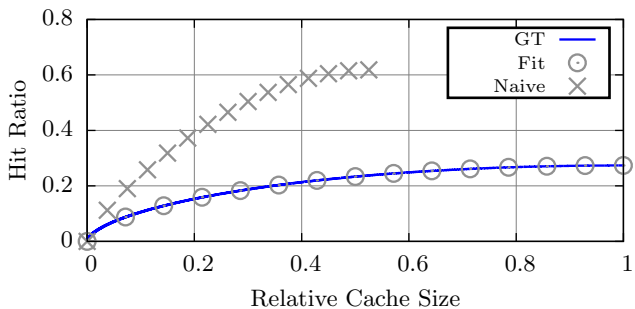}
  \end{subfigure}
  \caption{\textit{
      Hit ratio of a cache fed by
			\emph{\data{prt}} trace: ground-truth (GT) and prediction by the naive
      estimation. The cache size is normalized with respect to that of the GT}
  }
  \label{fig:hr_naive}
\end{figure}
In Figure~\ref{fig:hr_naive}, we show the impact of these limitations for the
hit ratio estimation, based on the \data{prt} trace.  The first curve is our
ground-truth. It is obtained via simulation of a LRU cache starting from an
empty cache; the cache is fed by the traffic trace that is randomly shuffled to
enforce the IRM assumption.  The second curve is the prediction of the IRM
model, when fed by the real popularities in the trace (see
Section~\ref{sec:irm_che} for a quick derivation of the transient hit ratio for
the IRM).  As expected, it perfectly fits the ground-truth. The third curve
shows the results obtained by the IRM model when fed by the parameters
$\Est{\CatSize}^{\nv}$ and
$\Est{\Pop}_\iDoc^{\nv},1\leq\iDoc\leq\Est{\CatSize}^{\nv}$, from the naive
estimation. The hit ratio curves are seen to clearly differ, and the naive
method proves inaccurate for estimating document popularities when fitting a
performance model.

In the absence of any prior knowledge about the popularity distribution, the
only available data for the estimation of each document popularity is a single
request count, which limits the accuracy of this approach.  To overcome this
lack of information, we thus aim at jointly estimating the set of popularities,
from the joint set of request counts. The latter approach allows us to use all
the information contained in the joint Poisson distribution rather than just
the mean.

Our problem can therefore be stated as follows:
\begin{ps}
Given the measured request counts $\lbrace \FSeq{\NReqsR}{\ObsDoc} \rbrace$,
determine the parameters $\Est{\CatSize}$ and $\Est{\Pop}_1$, $\Est{\Pop}_2$,
$\ldots, \Est{\Pop}_{\Est{\CatSize}}$ so that the set of random variables
$\lbrace \FSeq{\NReqs}{\Est{\CatSize}} \rbrace$, where $\NReqs_\iDoc =
\Poiss{\Est{\Pop}_\iDoc}$ for $1\leq\iDoc\leq\Est{\CatSize}$, is the
``closest'' to the set $\lbrace\FSeq{\NReqsR}{\ObsDoc}, 0, \ldots, 0\rbrace$,
with $\Est{\CatSize} - \ObsDoc$ zeros at the tail.
\end{ps}

\section{Maximum Likelihood Estimation}
\label{sec:est}

In this section, we show how to solve the latter inverse problem via the
Maximum Likelihood method.

In the IRM setting, the parameters $(\FSeq{\Pop}{\CatSize}, \CatSize)$ are not
ordered, and thus every request count could correspond to any of the
popularities. The likelihood given observations $\FSeq{\NReqsR}{\CatSize}$
thus runs through every permutation $\sigma$ of size $\CatSize$. Specifically
the likelihood is given by
\[
\rec{K!}
\sum_{\sigma}
\left(
	\prod_{j = 1}^{\CatSize_0}
\frac{e^{-\Pop_{\sigma(j)}} \Pop_{\sigma(j)}^{\NReqsR_j}}
     {\NReqsR_j!}
     \times
     \prod_{j=\CatSize_0 + 1}^{\CatSize}
     e^{-\Pop_{\sigma(j)}}
     \right).
\]
This combinatorial explosion for large $K$ makes the ML method intractable for
the IRM model. We thus propose in the following a slightly modified model,
which is simultaneously tractable for ML estimations and simple to analyze.

\subsection{IRM Mixture Model (IRM-M)}
\label{sec:est_mixt}

In order to succinctly describe the popularity parameters
$\FSeq{\Pop}{\CatSize}$ and to ease their estimation, we slightly modify the
IRM model by considering them as random variables.
Specifically, we assume that $\FSeq{\Pop}{\CatSize}$ are an \iid sample from an
unknown \emph{mixing distribution} with density $\dens$. Given the value of
$\Pop_\iDoc$, the request process to the $\iDoc\th$ document remains a Poisson
process of intensity $\Pop_\iDoc$, and thus the counts of each document follow
a mixed Poisson distribution with mixing distribution $\dens$. In particular,
the number of requests $\NReqs$ for any document satisfies
\begin{align}
	\label{eq:proba_mixed}
	\PP{\NReqs = j} &= \E{\frac{e^{-\Pop}
            \Pop^\iCount}{\iCount!}}[][][\dens]  =
        \int_{0}^\infty \frac{e^{-\Pop} \Pop^\iCount}{\iCount!}
        \dens(\Pop) d\Pop
\end{align}
for $j\in\N$, where the operator $\E{\cdot}[][][\dens]$ represents the
expectation under the mixing distribution $\dens$.

\begin{figure*}[!t]
	\begin{subfigure}{0.5\textwidth}
		\label{fig:delta_mixing_np}
		\centering
		\input{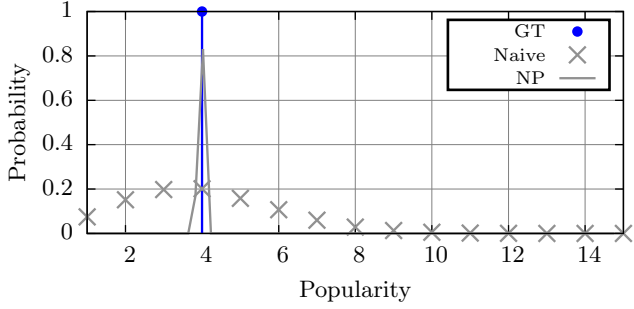}
	\end{subfigure}
	\begin{subfigure}{0.5\textwidth}
		\label{fig:pareto_mixing}
		\centering
		\input{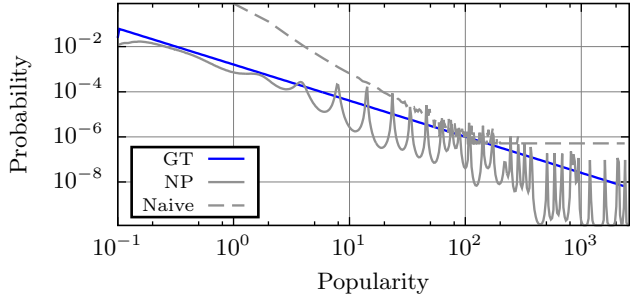}
	\end{subfigure}
	\caption{\textit{
			Mixing distribution obtained via the non-parametric
			methods for the \emph{\data{delta}} (left) and \emph{\data{prt}}
            (right) traces}
	}
	\label{fig:mixing}
\end{figure*}

\subsection{ML estimation on IRM-M}
By modifying the model, we have changed the problem of estimating the static
parameters $\FSeq{\Pop}{\CatSize}$, to that of estimating the mixing
distribution $\dens$.
\begin{ps}[IRM-M]
Given the measured request counts $\lbrace \FSeq{\NReqsR}{\ObsDoc} \rbrace$,
determine the catalog size $\Est{\CatSize}$ and the mixing density
$\Est{\dens}$ such that an \iid mixed Poisson sample $\lbrace
\FSeq{\NReqs}{\Est{\CatSize}} \rbrace$ is the ``closest'' to the set \\
$\lbrace \FSeq{\NReqsR}{\ObsDoc}, 0, \ldots, 0\rbrace$, with $\Est{\CatSize} -
\ObsDoc$ zeros at the tail.
\end{ps}

We now show how this problem can be solved via a ML method. Let $\MaxCount =
\max_{\iDoc=1}^{\ObsDoc} \lbrace \NReqsR_\iDoc \rbrace$  be the maximum number of requests
over all documents, and let
\[
	\EmpMeas_\iCount
	=
	\rec{\ObsDoc}
	\sum_{\iDoc=1}^{\ObsDoc} \II{\NReqsR_\iDoc = \iCount}
\]
be the proportion of documents with $\iCount$ requests, $1 \! \! \! \leq \! \! \! \iCount \! \! \! \leq \! \! \! \MaxCount$.
Using \eqref{eq:proba_mixed},
the log-likelihood $\lik{\dens}{\EmpMeas}$ of the popularity distribution
$\dens$ for the observations $\EmpMeas=\left(\EmpMeas_\iCount\right)_{j\geq 1}$
reads as
\begin{align*}
	\lik{\dens}{\EmpMeas}
	&=
	\sum_{\iCount=1}^\MaxCount
	\EmpMeas_\iCount \log \PP{\NReqs = \iCount}[][][\NReqs > 0]
	\\
	&=
	\sum_{\iCount=1}^\MaxCount
	\EmpMeas_\iCount \log \E{\frac{e^{-\Pop} \Pop^\iCount}{\iCount!}}[][][\dens]
	-
	\log \E{1 - e^{-\Pop}}[][][\dens] .
\end{align*}
We remark that in this setting, the catalog size $\CatSize$ is decoupled from
the popularity distribution. Thus, we can first obtain an estimator
$\Est{\dens}$ of the mixing distribution $\dens$, and then approximate
$\CatSize$ by
\begin{equation}
\label{eq:catsize_est}
	\Est{\CatSize}^{\ml} = \frac{\ObsDoc}{\E{1-e^{-\Pop}}[][][\Est{\dens}]}
\end{equation}
which is asymptotically close to the ML estimator.

We now proceed with the detailed form of the likelihood function for the
\emph{parametric} and \emph{non-parametric} estimation procedures.  In both
approaches, we numerically solve the problems with a generic non-linear
optimization solver in MATLAB based on an interior point algorithm. Our code is
freely available online.\footnote
{Code : \url{http://www.olmos.cl/code/mixed_poisson.tgz}} We
discuss the use of specialized algorithms in Section~\ref{sec:disc}.

\subsubsection{Parametric Estimation}
In this setting, we determine the mixing distribution within a parametric
family of density functions. The choice of that parametric family relies on an
a-priori knowledge. The computation of the ML estimator obviously depends on
this choice, and due to space restriction, we here limit ourselves  to the
two-parameter Pareto family
with densities
$
	\dens(x) =  \alpha x_m^\alpha / x^{\alpha + 1}
$
for $x > x_m$, with $\alpha$, $x_m$ the shape and scale parameters,
respectively.  The log-likelihood of parameters $\alpha$ and $x_m$ then reads
\begin{align*}
	\lik{\alpha, x_m}{\EmpMeas}
	=&
	\sum_{\iCount=1}^\MaxCount
	\EmpMeas_\iCount \log \frac{\Gamma(j-\alpha, x_m)}{j!}
	\,
	-\log \left( \alpha x_m^\alpha - \Gamma(-\alpha, x_m)
        \right).
\end{align*}

\subsubsection{Non-Parametric Family}
In the absence of a-priori knowledge about the distribution $\dens$, the
non-parametric (NP) approach provides a method to obtain an estimator.
In this setting, we determine a discrete distribution $\dens$ of the form
$
	\PP{\Pop = x_\iPar} = \Par_\iPar$
for $1 < \iPar < \NPar$. The log-likelihood correspondingly reads
\[
	\lik{\Par}{\EmpMeas}
	=
	\sum_{\iCount=1}^\MaxCount
	\EmpMeas_\iCount
	\log
	\sum_{\iPar=1}^\NPar \Par_\iPar \frac{e^{-x_\iPar} x_\iPar^\iCount}{\iCount!}
	-
	\log \sum_{\iPar=1}^\NPar \Par_\iPar (1 - e^{-x_\iPar}) .
\]
\subsection{Hit Ratio Analysis}
As detailed in the Appendix~\ref{sec:appendix}, the IRM-M model
proves to be tractable for evaluating the performance of an LRU cache.
In particular, the so-called ``Che approximation'' is easily adapted to
the IRM-M case;
furthermore, we are able to derive formulas for the transient analysis of the
hit ratio, when starting from an empty cache.

\section{Numerical Evaluation}
\label{sec:apps}
The accuracy of the parameter estimation can be evaluated at three different
levels, as expressed by the following questions: \textbf{(1)} Is the estimated
popularity density close to the actual popularity density? \textbf{(2)} Is the
request flow predicted by the model statistically similar to the actual request
flow? \textbf{(3)} Is the performance indicator of the fitted model, e.g., the
hit ratio, accurately predicted?

Throughout this section, we assess the precision of a curve
estimate by computing the so-called
\emph{mean absolute relative error} (MARE). More precisely, the MARE
between a reference sequence of points (or curve)
$(x_i)_{1\leq i \leq N}$ and an estimate sequence
$(y_i)_{1\leq i \leq N}$ is defined by
\[\text{MARE}(X,Y)= \frac{1}{N}\sum_{i=1}^N \frac{|y_i-x_i|}{|x_i|}.\]

\subsection{Estimation of popularity distribution}
\label{sec:apps_popu}
First, we start with the most general question, that is, the estimation of the
mixing distribution. Such an inverse problem is known to be ill-posed.

For the NP estimation, we obtain an estimate $\Est{\dens}^{\np}$ of the
popularity density by applying the NP method, using a support with $0.01$ as
lower bound, exponentially increasing spacings and an upper bound slightly
larger than the maximum of observed requests (e.g., $2\,400$ for \data{prt} and
$16$ for \data{delta}). The naive fitting corresponds to the empirical measure
of the request counts, that is, the mixture of Dirac measures
$\rec{\EffCatSize} \sum_{\iDoc=1}^{\EffCatSize}\delta_{\NReqsR_\iDoc}(.)$.

We observe in Figure~\ref{fig:mixing} the NP estimator of the mixing
distribution for the \data{delta} and \data{prt} datasets. In the \data{delta}
case, the ground-truth is a Dirac measure at $\Pop=4$, and the naive method
fails at correctly estimating its shape, whereas the ML estimator concentrates
its mass around the value $\Pop=4$. In the \data{prt} case, the estimated
distribution is irregular, tending to accumulate mass at certain points (see
Section~\ref{sec:disc_inference} for possible regularization solutions). This
concentration is no surprise, since in the non-censored case the ML estimator
is discrete probability distribution~\cite{lindsay1995mixture}. The peaks,
nevertheless, capture the power law trend, as reflected by the good estimation
quality of the mixture distribution. In contrast, the naive method fails at
correctly estimating both the trend of distribution body and its tail.

\begin{figure}[!tb]
	\centering
	\input{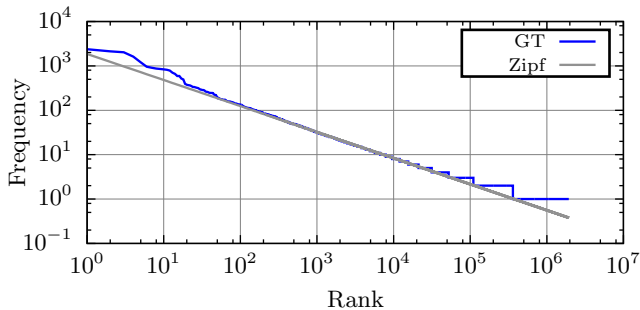}
	\caption{
     \textit{Rank frequency distribution for the \emph{\data{prt}} trace}}
    \label{fig:rank_freq_fit}
\end{figure}

Using Equation~\eqref{eq:catsize_est}, we also calculate the catalog size,
giving $\Est{\CatSize} \approx 11\,600\,000$ (resp. $105\,278$) for the
\data{prt} (resp. \data{delta}) case. This represents a relative error of
$11.6\%$ and $5.2\%$, respectively. Following Equation~\eqref{eq:catsize_est},
it shows that estimating the probability that a document receives no request
for the duration of the trace, based on the very same trace, is a difficult
task. As a consequence, this error is not negligible. It is, however, smaller,
and even more significantly in the \data{prt} case, than the relative error of
the naive method (recall that $\Est{\CatSize}^{\nv} = \ObsDoc = 1\,900\,000$
and $\Est{\CatSize}^{\nv} = 92\,046$ for the \data{prt} and \data{delta}
traces, respectively).

When some a priori knowledge about the distribution shape is available, the
estimates can be improved via the parametric approach. In the \data{prt} case,
the resulting Pareto fit gives the estimates $\Est{\alpha} = 1.597$ and
$\Est{x}_m = 0.099$ that are very close to the original parameters $\alpha=1.6$
and $x_m = 0.1$. We compare these results to that of the ``log-log'' approach,
which consists in estimating the tail index by fitting a least square
approximation to the log-log rank-frequency plot, as shown in
Figure~\ref{fig:rank_freq_fit}. The rank frequency plot roughly decays as
$1/\alpha$. Using the first $20\,000$ objects to compute the regression, the
estimation gives $1.704$, which is worse than the ML estimate.

\begin{figure*}[!th]
 \begin{subfigure}{0.33\textwidth}
  \centering
  \caption{\textbf{\emph{\data{prt}}: First ten points}}
  \input{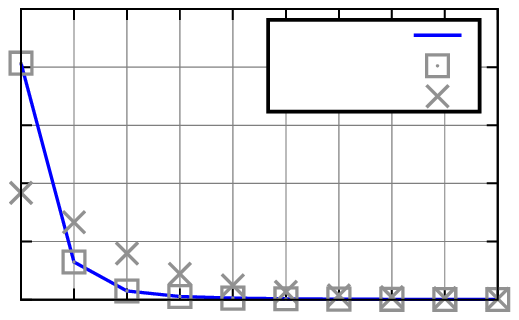}
  \label{fig:pareto_mixed_head}
 \end{subfigure}
 \begin{subfigure}{0.31\textwidth}
  \centering
  \caption{\textbf{\emph{\data{yt}}: First ten points}}
  \input{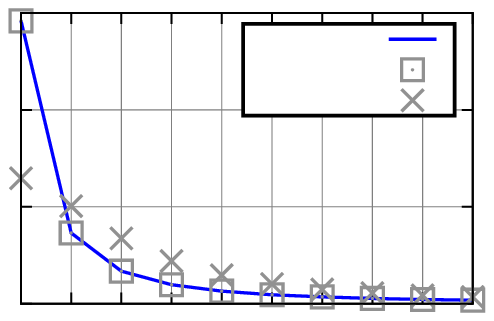}
  \label{fig:otu_mixed_head}
 \end{subfigure}
 \begin{subfigure}{0.31\textwidth}
  \centering
  \caption{\textbf{\emph{\data{vod}}: First ten points}}
  \input{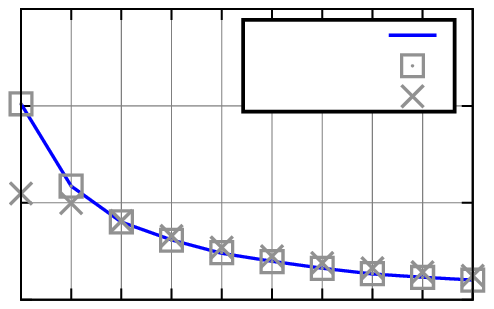}
  \label{fig:vod_mixed_head}
 \end{subfigure}
 \hfill\\
 \begin{subfigure}{0.35\textwidth}
  \centering
  \caption{\textbf{\emph{\data{prt}}: Full range in log scale}}
  \input{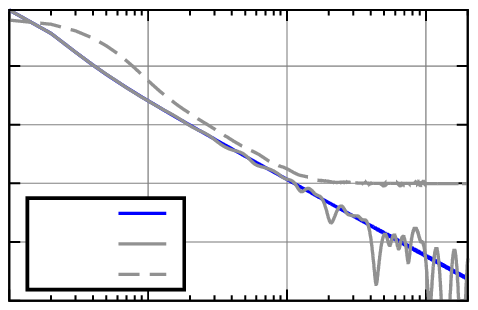}
  \label{fig:pareto_mixed_full}
 \end{subfigure}
 \begin{subfigure}{0.32\textwidth}
  \centering
  \caption{\textbf{\emph{\data{yt}}: Full range in log scale}}
  \input{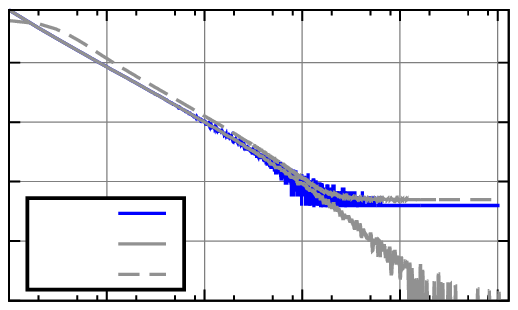}
  \label{fig:otu_mixed_full}
 \end{subfigure}
 \begin{subfigure}{0.32\textwidth}
  \centering
  \caption{\textbf{\emph{\data{vod}}: Full range in log scale}}
  \input{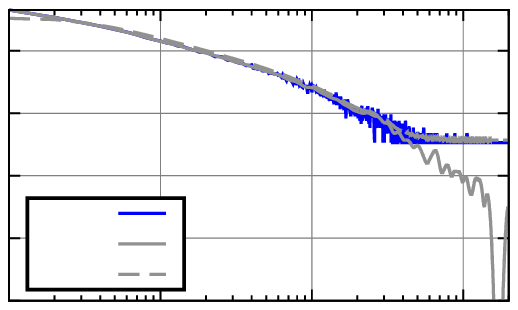}
  \label{fig:vod_mixed_full}
 \end{subfigure}
 \hfill
 \caption{
  \textit{Censored Mixture distribution estimations obtained with the non-parametric method}
 }
 \label{fig:mixed}
\end{figure*}

\subsection{Request flow estimation}
In this section, we specify the discussion by estimating the zero-censored
request count distribution (or mixture distribution in statistical terms)
$\PP{\NReqs = j}[][][\NReqs > 0]$, $j \geq 1$.

For the naive approach, we generate $50\,000$ IRM traces using the estimated
parameters.  We then calculate the average empirical distribution of the
request per document.  The number of generated traces ensures a coefficient of
variation lower that $10^{-4}$ for all points of the distribution. As regards
the ML approach, using the $\Est{\dens}^{\np}$ density, we compute
the associated zero-censored request distribution using~\eqref{eq:proba_mixed}.

In Figure~\ref{fig:mixed}, we show the resulting zero-censored request
distribution estimated by each method. For comparison, we include the real
mixture distribution for the \data{prt} dataset, which can be calculated
explicitly. For the \data{yt} and \data{vod} datasets, we show instead the
observed request distribution.

Two issues are raised by the naive approach, that are not present in the
maximum likelihood estimation:\\
-- first, at the head of the distribution,
where most of the mass is concentrated, large estimation errors are produced by
the naive approach. Such errors produce a mass shift towards the tail of the
distribution. On the contrary, the NP estimation matches perfectly the head of
the distribution;\\
-- second, the naive method over-fits the tail of the
distribution. We observe in Figure~\ref{fig:pareto_mixed_full} that the naive
estimate shows a ``horizontal branch'' at the tail, and differs significantly
from the ground-truth that is approximately a straight ``diagonal'' line. This
horizontal branch is in fact a few isolated masses, though they look as a line
on the figure. The naive estimation therefore concentrates the mass of the
ground-truth distribution on a few points. On the other side, the ML
estimation correctly estimates the trend of the distribution at all scales,
though noise inaccuracies appear at the tail. This is quantified by the MARE
of 1.67 for the ML estimation, whereas the naive method leads to a MARE of 668,
for the full range distribution.
As regards the \data{yt} and \data{vod} cases in
Figures~\ref{fig:otu_mixed_full} and~\ref{fig:vod_mixed_full}, we similarly
observe the same horizontal branch at the tail for the naive distribution. In
the absence of available ground-truth, we do not compute the MARE, but the
similarity of behavior hints that the ML method also performs better on these
traces.

\subsection{Hit Ratio Estimation}
We finally compare the hit ratios predicted by the IRM-M model with popularity
distributions fitted using the naive and the ML methods, both for the
\data{prt} and \data{yt} traces.

\begin{figure}[t!]
	\centering
	\input{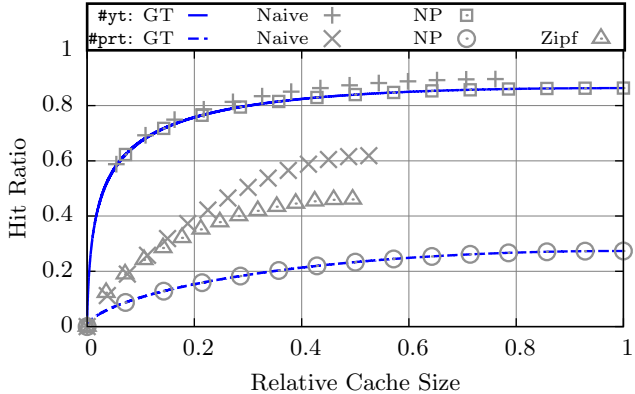}
	\caption{\textit{
			Hit ratio for \emph{\data{yt}} and \emph{\data{prt}} datasets.
		The cache size is normalized with respect to
		that of the ground-truth in the \emph{\data{prt}} case and with respect to
		$\Est{\CatSize}^{\ml}$ in the \emph{\data{yt}} case.
	}}
	\label{fig:hr_nonpar}
\end{figure}

Figure~\ref{fig:hr_nonpar} shows the obtained hit ratio curve in each case. The
ground-truth curves are obtained by simulation of a LRU cache fed by the
shuffled traces. The Naive (resp. NP) curves are obtained when using
Formula~\eqref{eq:hr_irm_trans} (resp. \eqref{eq:hr_irmm_trans}) with the
parameters obtained by the naive (resp. NP) method. Finally, the Zipf curve,
for the \data{prt} trace, corresponds to the hit ratio prediction when using
the ``log-log'' parametric fitting method detailed in
Section~\ref{sec:apps_popu}.

The naive approach leads to small inaccuracy for the \data{yt} trace and large
errors for the \data{prt} trace, with respective MARE of 0.06 and 1.44. This
difference in estimation accuracy can be explained by the variability of the
random variable $\NReqs$. Indeed, in the \data{yt} dataset, documents receive
an average of 7.3 requests per document, whereas this average decreases to
$1.4$ in the \data{prt} trace. It follows that the coefficient of variation of
the request count distribution is greater in the \data{prt} trace than in the
\data{yt} trace. As expected, the inaccuracy of the naive method is greater for
the former than for the latter. Note also that from an operational point of
view, the focus is on the miss ratio, which determines the dimensioning
requirements upstream of the cache. The inaccuracy of the naive hit ratio
prediction for the \data{yt} dataset becomes relatively significant in this
context. As shown by the Zipf curve, the knowledge of a relevant parametric
family allows us to improve the hit-ratio estimation. The error, however,
remains significant with a MARE of 0.96. In contrast, the non-parametric ML
curves match perfectly the original ones, as shown by the MARE of 0.002 for the
\data{yt} trace and 0.005 for the \data{prt} trace. We conclude that, as
regards hit ratio, our estimation method accurately estimates the model
parameters. In contrast, in the Zipf case, a seemingly small error of 0.1 in
the estimation of the tail exponent leads to a significant error in the hit
ratio estimation.

\label{sec:apps_hr}

\section{Discussion and Conclusion}
\label{sec:disc}

\subsection{Extendability to other models}
\label{sec:extension}
Our method relies exclusively on the fact that we explicitly know the request
count distribution of a document, given its popularity. As a consequence, our
framework can be extended to other traffic models such as renewal
traffic~\cite{fofack2012analysis, berger2014TTL} and the \emph{Shot-Noise} (SN)
traffic~\cite{olmos2014catalog, traverso2013temporal}, though the formulation
of the ML function changes in each case.

In practice, however, this reformulation introduces new challenges to the
inverse problem. First, note that in the IRM-M case, the requests processes are
particularly tractable for the inverse problem, because they are entirely
characterized by a single parameter and exhibit no correlation.  However, for
other processes, the fitting of the point processes may require to fit a
multivariate distribution \cite{olmos2014catalog} or a whole random
distribution \cite{fofack2012analysis} per document.
Second, due to the greater importance of the time variable in other models, the
censoring effects coming from the finite observation window are more severe.
Additional care must be taken when adapting our estimation method to other
models.

\subsection{Maximization techniques}
\label{sec:disc_inference}

The main current limitation of our maximization approach is that the estimated
mixing density exhibits a lot of peaks, which is consistent with the results of
Lindsay~\cite{lindsay1995mixture}. This might be a problem when one aims at
understanding the nature of the popularity distribution.

A possible solution to enforce smoothness in the mixing density estimation is
to introduce a penalization for the irregularities. Classical candidates for
penalization are the $L^2$-penalization or a logarithmic penalization
$ R(\Par) = \sum_{\iPar=1}^\NPar (\Par_{\iPar+1} -
  \Par_\iPar)(\log \Par_{\iPar+1} - \log \Par_\iPar) /
  (\Supp_{\iPar+1} - \Supp_\iPar)$.
One then  maximizes $\lik{\Par}{\EmpMeas} - \SmoothPar R(\Par)$, where
$\SmoothPar$ represents the trade-off factor between fitness and smoothness.
Regularization here comes at the price of choosing the right penalization
function $R(\cdot)$ and the right value of $\SmoothPar$ and in our case, the
results have been satisfactory only for concentrated mixing distributions.

Another possibility is to exploit the fact that the peaks conserve the overall
trend of the distribution.  We thus extract the peak locations. A second ML
optimization is then performed using these peak locations as the new support.
Though non-standard, this gives satisfactory results for the \data{prt} dataset
(not shown here due to lack of space).

\subsection{Summary of results}

In this paper, we have presented and solved the inverse problem that consists
in estimating from a trace the  popularity parameters to be used in a
performance model. A key point in our approach is that we consider the
probability that a document receives  a given number of requests, rather than
the probability that a request is directed to a given document. This
representation is consistent with recently developed caching models
\cite{olmos2014catalog,traverso2013temporal,fofack2012analysis}. Moreover, it
allows us to avoid the fitting of  a rank-frequency plot, which is in essence
an order statistics and may exhibit undesirable properties. Our second
contribution on the modeling aspects is that we consider popularities as random
variables, rather than  parameters, leading to a tractable mixture model.

The inverse problem stems from the random nature of the requests count $\NReqs$
for a given document.  In particular, a traffic trace contains a single sample
of these requests counts. The accuracy of any method that aims at fitting
independently the popularity of each document is therefore limited by the
inherent variability of the random variable $\NReqs$.  The importance of using
a sound methodology correspondingly increases when the variability of the
request counts is large, which is typically the case when $\NReqs$ is small.

Determining the parameters of the model allows one to use the performance for
several objectives, including the dimensioning of operational networks or the
design of new mechanisms. More importantly, in contrast with simulation-based
analysis, it enables one to more easily explore ``what-if'' scenarios, by
keeping some parameters at their current value and modifying others to reflect
future or possible changes.

\section{Appendix}
We here detail the derivation of hit ratio formulas for the IRM-M
model.
\label{sec:appendix}
\subsection{IRM Model}
\label{sec:irm_che}
For comprehension purposes, we first briefly review the ``Che approximation''
method for the hit ratio estimation in the IRM model (additional details can be
found in~\cite{fricker2012versatile}).  Given popularities
$\FSeq{\Pop}{\CatSize}$, let $X^\iDoc(t)$ denote  the number of different
documents, apart from the $k$-th, requested in a time window $[0,t]$, that is,
\[
	X^\iDoc(t)
	=
	\sum_{i=1; i \neq \iDoc}^\CatSize \II{\NReqs_i[0,t] \geq 1}.
\]
Let
$
	\ExitTime{\Csize}^{\iDoc}
	=
	\inf \lbrace t > 0: X^\iDoc(t) \geq \Csize \rbrace
$
be the exit time to level $\Csize$  for process $X^\iDoc$;
$\ExitTime{\Csize}^{\iDoc}$ represents the eviction time for content $\iDoc$ in
a LRU cache of size $\Csize$, given that it is not requested during this time
period.  Now, the core of the ``Che approximation'' consists in the two
following steps:
\begin{compactenum}
\item all $\ExitTime{\Csize}^\iDoc$ have the same
  distribution, i.e., $\forall k, \ExitTime{\Csize}^\iDoc \distEqual \ExitTime{\Csize}$;
\item the random variable $\ExitTime{\Csize}$ is well approximated by a constant $\CharTime{\Csize}$ called the ``characteristic time''. The time $\CharTime{\Csize}$
  is  implicitly defined by the equation
  \begin{equation}
		\sum_{\iDoc=1}^\CatSize
		\E{ \II{\NReqs_\iDoc[0, \CharTime{\Csize}] \geq 1}}
		=
    \sum_{\iDoc=1}^\CatSize
		1 - e^{-\Pop_\iDoc\CharTime{\Csize}}
    =
    \Csize.
    \label{eq:char_time_def}
	\end{equation}
\end{compactenum}
Intuitively, $\CharTime{\Csize}$ is the time when, on average, $\Csize$
different objects have been requested.

In the stationary case, the hit ratio $\HitRatio$ can then be derived as follows.
Using the 
PASTA property, the hit ratio of document $\iDoc$ for a cache of
size $\Csize$ is equal to
$1-e^{\Pop_\iDoc \CharTime{\Csize}}$, and by averaging on all documents, it follows that
\begin{equation}
\HitRatio \approx \rec{\Lambda} \sum_{\iDoc=1}^\CatSize \Pop_\iDoc (1 -
e^{-\Pop_\iDoc \CharTime{\Csize}})\,.
\label{eq:hr_irm_statio}
\end{equation}

In the transient case, we simply assume that
$\ExitTime{\Csize}^\iDoc \leq \WinSize$ (the hit ratio does not increase with
$\ExitTime{\Csize}^\iDoc$ when $\ExitTime{\Csize}^\iDoc > \WinSize$).  By
independence, it can be shown (see Proposition 3, \cite{olmos2014catalog}) that
the average number of hits for the $\iDoc$-th document in a time window of size
$\WinSize$, starting from an empty cache, is
$\E{\AveNHits(\Pop_\iDoc,\ExitTime{\Csize}^\iDoc) }$ where the expectation
carries on $\ExitTime{\Csize}^\iDoc$ and the function $\AveNHits(\Pop,t)$ is
defined by
\begin{equation}
	\AveNHits(\Pop,t)
	=
		(\Pop W - 1)(1- e^{-\Pop t}) + \Pop t e^{-\Pop t}
	,\quad t < \WinSize .
        \label{eq:h_def}
\end{equation}
In consequence, setting
$\Lambda = \sum_{\iDoc=1}^\CatSize \Pop_\iDoc$,
the transient hit ratio $\HitRatio(\WinSize)$ is given by
\[
  \HitRatio(\WinSize)
  =
  \rec{\Lambda \WinSize}
  \sum_{\iDoc=1}^{\CatSize} \E{\AveNHits(\Pop_\iDoc,\ExitTime{\Csize}^\iDoc)}.
\]
Applying the ``Che approximation'', we then obtain
\begin{equation}
  \HitRatio(\WinSize)
   \approx
     \rec{\Lambda} \sum_{\iDoc=1}^\CatSize \Pop_\iDoc (1 - e^{-\Pop \CharTime{\Csize}})
     + \rec{\Lambda \WinSize}
     \left(
     \sum_{\iDoc=1}^\CatSize \Pop_\iDoc\CharTime{\Csize} e^{-\Pop_\iDoc \CharTime{\Csize} } - \Csize
     \right)
   \label{eq:hr_irm_trans}.
\end{equation}
The second term of~\eqref{eq:hr_irm_trans} vanishes as $\WinSize \to \infty$,
leading to equality~\eqref{eq:hr_irm_statio} for the stationary hit
ratio.

\subsection{IRM-M Model}
\label{sec:irmm_che}
We now address the IRM-M case. We first show how to derive the hit ratio in
this setting; we further prove formally the validity of the ``Che
approximation'' in the case where $\Csize=\delta\CatSize$ and $\CatSize$ tends
to infinity.

$\bullet$ Given the popularities $\FSeq{\Pop}{\CatSize}$, let us define $X^\iDoc$,
$\ExitTime{\Csize}^\iDoc$ as in the previous section, and let
$\delta=\Csize/\CatSize$ be the proportion of stored documents.
As the popularities are here an \iid sample, and since $X^\iDoc$ and
$\ExitTime{\Csize}^\iDoc$ are independent of $\Pop_\iDoc$, the previous
quantities do not consequently depend on the document index $\iDoc$. In
consequence, this validates the first step of the ``Che approximation''.

For the second step, define the characteristic
time $\CharTime{\delta}$ as
\begin{align}
\label{eq:irmm_che}
\CharTime{\delta}=r^{-1}\left(\delta\right)
\quad
\text{ with } \quad
r(t) &= \E{1 - e^{-\Pop t}},
\end{align}
which is  equivalent to dividing
both sides of~\eqref{eq:char_time_def} by $\CatSize$.
Following the same steps as in the previous section, it is easy to derive the
following hit ratio formulas:
\begin{align}
\HitRatio &\approx
\frac{\E{\Pop (1-e^{-\Pop\CharTime{\delta}})}}{\E{\Pop}}
,
\label{eq:hr_irmm_statio}
\\
\HitRatio(\WinSize) &\approx
\frac{\E{\Pop (1-e^{-\Pop \CharTime{\delta}})}}{\E{\Pop}}+
\frac{\E{\Pop\CharTime{\delta}e^{-\Pop \CharTime{\delta}}}- \delta}{\E{\Pop}W}.
\label{eq:hr_irmm_trans}
\end{align}
Equations~\eqref{eq:hr_irmm_statio}
and \eqref{eq:hr_irmm_trans} are the
IRM-M analogs of~\eqref{eq:hr_irm_statio}
 and \eqref{eq:hr_irm_trans}.

$\bullet$ We show that the second step of the Che
approximation is asymptotically exact, that is, the random
variable $\ExitTime{\Csize}$ can be replaced by the associated
characteristic time $\CharTime{\delta}$.
Consider the case where the cache size scales with the catalog size, that is,
$\delta$ remains constant, and $\Csize$ and $\CatSize$ grow to infinity.
Recall that the distribution of $\ExitTime{\Csize}$ is given by
\[
	\PP{\ExitTime{\Csize} > t}
	=
	\PP{\sum_{\iDoc=1}^\CatSize \II{\NReqs_\iDoc[0,t] \geq 1} < \Csize}
\]
for $t \geq 0$, which can be rewritten as
\begin{equation}
	\label{eq:exit_time_scale}
	\PP{\ExitTime{\delta \CatSize} > t}
	=
	\PP{\rec{\CatSize} \sum_{\iDoc=1}^\CatSize  \II{\NReqs_\iDoc[0,t] \geq 1} < \delta}
	.
\end{equation}
An application of the law of large numbers shows that
\[
\lim_{K \to \infty}\rec{\CatSize} \sum_{\iDoc=1}^\CatSize  \II{\NReqs_\iDoc[0,t] \geq 1}
= r(t)
\]
almost surely; using~\eqref{eq:exit_time_scale}, $\ExitTime{\delta
	\CatSize}$ thus converges in probability to the constant $\CharTime{\delta}$, for
$\delta \in [0,r(\WinSize)]$, where $r(\WinSize)  = \E{\EffCatSize}/\CatSize$.
By the conditioning argument of Proposition 3 in~\cite{olmos2014catalog}, it can
be shown that the expectation of the number of hits $\NHits_\Csize = \NHits_{\delta \CatSize}$ 
satisfies the identity
$
\label{eq:ave_nmiss_exit_time}
\E{\NHits_{\delta \CatSize}}
	=
	\E{\AveNHits(\Pop,\ExitTime{\delta \CatSize})};
$
applying then the bounded convergence theorem (Section
13.6,~\cite{williams1991probability}) to the latter identity and dividing by
the expected number of requests $\E{\Pop}$ leads to
formulas~\eqref{eq:hr_irmm_statio} and~\eqref{eq:hr_irmm_trans}, as claimed.

\bibliographystyle{abbrv}
\bibliography{bibliography}

\begin{thebibliography}{10}

\bibitem{berger2014TTL}
D.~S. Berger, P.~Gland, S.~Singla, and F.~Ciucu.
\newblock Exact analysis of {TTL} cache networks: The case of caching policies
  driven by stopping times.
\newblock {\em arXiv preprint arXiv:1402.5987}, 2014.

\bibitem{carlinet2012four}
Y.~Carlinet, T.~D. Huynh, B.~Kauffmann, F.~Mathieu, L.~Noirie, and S.~Tixeuil.
\newblock {Four Months in DailyMotion: Dissecting User Video Requests}.
\newblock In {\em Wireless Communications and Mobile Computing Conference
  (IWCMC), 2012 8th International}, pages 613--618. IEEE, 2012.

\bibitem{cha2007tube}
M.~Cha, H.~Kwak, P.~Rodriguez, Y.-Y. Ahn, and S.~Moon.
\newblock I tube, you tube, everybody tubes: Analyzing the world's largest user
  generated content video system.
\newblock In {\em 7th ACM SIGCOMM conference on Internet measurement (IMC)},
  pages 1--14. ACM, 2007.

\bibitem{clauset2009power}
A.~Clauset, C.~R. Shalizi, and M.~E. Newman.
\newblock Power-law distributions in empirical data.
\newblock {\em SIAM review}, 51(4):661--703, 2009.

\bibitem{comte2015laguerre}
F.~Comte, V.~Genon-Catalot, et~al.
\newblock {Adaptive Laguerre density estimation for mixed Poisson models}.
\newblock {\em Electronic Journal of Statistics}, 9:1113--1149, 2015.

\bibitem{fofack2012analysis}
N.~C. Fofack, P.~Nain, G.~Neglia, and D.~Towsley.
\newblock Analysis of {TTL}-based cache networks.
\newblock In {\em 6th International Conference on Performance Evaluation
  Methodologies and Tools (VALUETOOLS)}, pages 1--10. IEEE, 2012.

\bibitem{fricker2012versatile}
C.~Fricker, P.~Robert, and J.~Roberts.
\newblock A versatile and accurate approximation for cache performance.
\newblock In {\em 24th International Teletraffic Congress}. IEEE Communications
  Society, 2012.

\bibitem{fricker2012impact}
C.~Fricker, P.~Robert, J.~Roberts, and N.~Sbihi.
\newblock Impact of traffic mix on caching performance in a content-centric
  network.
\newblock In {\em Computer Communications Workshops (INFOCOM WKSHPS), 2012 IEEE
  Conference on}, pages 310--315. IEEE, 2012.

\bibitem{gill2007youtube}
P.~Gill, M.~Arlitt, Z.~Li, and A.~Mahanti.
\newblock Youtube traffic characterization: A view from the edge.
\newblock In {\em 7th ACM SIGCOMM conference on Internet measurement (IMC)},
  pages 15--28. ACM, 2007.

\bibitem{gong2001tails}
W.~Gong, Y.~Liu, V.~Misra, and D.~Towsley.
\newblock On the tails of web file size distributions.
\newblock In {\em Proceedings of the annual Allerton Conference on
  Communucation Control and Computing}, volume~39, pages 192--201, 2001.

\bibitem{guillemin2013experimental}
F.~Guillemin, B.~Kauffmann, S.~Moteau, and A.~Simonian.
\newblock Experimental analysis of caching efficiency for youtube traffic in an
  {ISP} network.
\newblock In {\em 25th International Teletraffic Congress (ITC25)}, pages 1--9.
  IEEE, 2013.

\bibitem{guo2008stretched}
L.~Guo, E.~Tan, S.~Chen, Z.~Xiao, and X.~Zhang.
\newblock The stretched exponential distribution of internet media access
  patterns.
\newblock In {\em Twenty-seventh ACM symposium on Principles of distributed
  computing}, pages 283--294. ACM, 2008.

\bibitem{imbrenda2014icn}
C.~Imbrenda, L.~Muscariello, and D.~Rossi.
\newblock Analyzing cacheable traffic in {ISP} access networks for micro {CDN}
  applications via content-centric networking.
\newblock In {\em Proceedings of the 1st International Conference on
  Information-centric Networking}, ICN '14, pages 57--66. ACM, 2014.

\bibitem{karlis2001general}
D.~Karlis.
\newblock {A General EM Approach for Maximum Likelihood Estimation in Mixed
  Poisson Regression Models}.
\newblock {\em Statistical Modelling}, 1(4):305--318, 2001.

\bibitem{lindsay1995mixture}
B.~G. Lindsay.
\newblock {Mixture Models: Theory, Geometry, and Applications}.
\newblock {\em Institute for Mathematical Statistics: Hayward, CA}, 1995.

\bibitem{loiseau2009maximum}
P.~Loiseau, P.~Gon{\c{c}}alves, S.~Girard, F.~Forbes, and
  P.~Vicat-Blanc~Primet.
\newblock {Maximum Likelihood Estimation of the Flow Size Distribution Tail
  Index from Sampled Packet Data}.
\newblock In {\em ACM SIGMETRICS Performance Evaluation Review}, volume~37,
  pages 263--274. ACM, 2009.

\bibitem{olmos2014catalog}
F.~Olmos, B.~Kauffmann, A.~Simonian, and Y.~Carlinet.
\newblock Catalog dynamics: Impact of content publishing and perishing on the
  performance of a {LRU} cache.
\newblock In {\em 26th International Teletraffic Congress (ITC26)}, pages 1--9.
  IEEE, 2014.

\bibitem{oreshkin2014rate}
B.~N. Oreshkin, N.~Regnard, and P.~L'Ecuyer.
\newblock Rate-based daily arrival process models with application to call
  centers.
\newblock Technical report, 2014.

\bibitem{roberts2013exploring}
J.~Roberts and N.~Sbihi.
\newblock Exploring the memory-bandwidth tradeoff in an information-centric
  network.
\newblock In {\em 25th International Teletraffic Congress (ITC25)}, pages 1--9.
  IEEE, 2013.

\bibitem{roueff2005mixture}
F.~Roueff and T.~Rydn.
\newblock Nonparametric estimation of mixing densities for discrete
  distributions.
\newblock {\em The Annals of Statistics}, 33(5):2066--2108, 2005.

\bibitem{traverso2013temporal}
S.~Traverso, M.~Ahmed, M.~Garetto, P.~Giaccone, E.~Leonardi, and S.~Niccolini.
\newblock Temporal locality in today's content caching: Why it matters and how
  to model it.
\newblock {\em ACM SIGCOMM Computer Communication Review}, 43(5):5--12, 2013.

\bibitem{williams1991probability}
D.~Williams.
\newblock {\em {Probability with Martingales}}.
\newblock Cambridge University Press, 1991.

\end{thebibliography}

\end{document}